\documentstyle[twocolumn,aps,psfig]{revtex}
\begin{document}
\draft
\wideabs{          
\title{Quasiparticle thermal Hall angle and magnetoconductance in $\rm YBa_2Cu_3O_x$.} 
\author{K. Krishana, N. P. Ong, Y. Zhang, and Z. A. Xu$^\dagger$}
\address{Joseph Henry Laboratories of Physics, Princeton University, Princeton, New Jersey 
08544}
\author{R. Gagnon and L. Taillefer$^\S$}      
\address{Department of Physics, McGill University, Montreal, Quebec, Canada}
\date{\today}      


\maketitle                   

\begin{abstract}
We present a way to extract the quasiparticle ($qp$) thermal conductivity $\kappa_e$ and 
mean-free-path in $\rm YBa_2Cu_3O_x$, using the thermal Hall effect and the 
magnetoconductance of $\kappa_e$.  The results compare well with heat capacity experiments.  
Moreover, we find a simple relation between the thermal Hall angle $\theta_Q$ and the $H$-
dependence of $\kappa_e$, as well as numerical {\em equality} between $\theta_Q$ and the 
electrical Hall angle.  The results also reveal an anomalously anisotropic scattering process 
in the normal state.
\end{abstract}
}
The quasiparticle excitations in cuprate superconductors display many unusual properties 
related to their long transport lifetimes and the existence of nodes in the gap function.  
These properties have been investigated by microwave and heat transport techniques.  Bonn 
{\em et al.} \cite{Bonn} uncovered a broad peak in the zero-field microwave conductivity in 
$\rm YBa_2Cu_3O_7$ (YBCO).  From measurements of the the zero-field thermal conductivity 
anomaly in untwinned YBCO, Yu {\em et al.} \cite{Yu} inferred that it arises entirely from 
the quasiparticles. Taillefer {\em et al.} \cite{Taillefer} measured values 
of $\kappa_{xx}$ below 0.1 K that are close to the predicted universal value $\kappa_e^{00}$ 
\cite{Graf}. 

The long quasiparticle lifetime also produces a large thermal Hall effect which has been 
investigated in YBCO \cite{Krish1}.  While the quasiparticles are solely responsible for the 
thermal Hall conductivity $\kappa_{xy}$, their contribution to the diagonal conductivity 
$\kappa_{xx}$ is harder to sort out experimentally. The general problem of $qp$ transport in 
a $d$-wave superconductor in a field is an active area of research 
\cite{Simon,Hirschfeld,Anderson,Franz}.  Novel effects such as 
those arising from the doppler-shift term \cite{Hirschfeld}, field quantization of the states 
around the nodes \cite{Anderson}, and Andreev scattering from a disordered vortex array 
\cite{Franz} have been proposed. 

We report results that reveal a close relation between $\kappa_{xy}$ and $\kappa_{xx}$ in 
YBCO.  We find that the thermal Hall angle $\theta_Q$ closely tracks a parameter $p(T)$ 
extracted from the field dependence of $\kappa_{xx}$, which implies strongly that $p(T)$ is 
proportional to the $qp$ mean-free-path ($mfp$).  In addition, we find that $\theta_Q$ agrees 
{\em numerically} with the electrical Hall angle $\theta_e$ extrapolated from above $T_c$.  
Our results disagree with the factor of 2 discrepancy found in a recent related study 
\cite{Zeini}.  

Measurements of $\kappa_{xx}$ and $\kappa_{xy}$ were made on 4 crystals (1-4) in a field $\bf 
H\parallel c$.  Samples 1 and 2 are both optimally doped and detwinned with $x$ = 6.95 and 
$T_c$ = 93.3 K.  With the thermal current ${\bf J}_Q \parallel \hat{\bf x}\parallel {\bf a}$, 
the thermal gradient $-\partial_x T$ is measured with a pair of matched cernox sensors.  The 
antisymmetric `Hall' gradient $-\partial_y T\parallel \bf b$ (the chain axis) is measured 
with a pair of chromel-constantan thermocouples, as in \cite{Krish1}.  Samples 3 and 4 are 
underdoped twinned crystals with $T_c$ = 63 and 60 K, respectively (in both $x$ = 6.63).

As our analysis rests on the finding that $\kappa_{ph}$ (the phonon conductivity) is 
insensitive to $H$, we first summarize the evidence for this result.  In the cuprates, the 
field dependence of $\kappa_{xx}$ is known to be described by the equation
\begin{equation}
\kappa_{xx}(H,T)= \frac{\kappa_e^0(T)}{(1+p(T)|H|^\mu)} + \kappa_B(T),
\label{kxx}
\end{equation}
where $p(T)$ is an inverse field scale and $\mu = 1$.  

\begin{figure} [h]
\centerline{ \psfig{figure=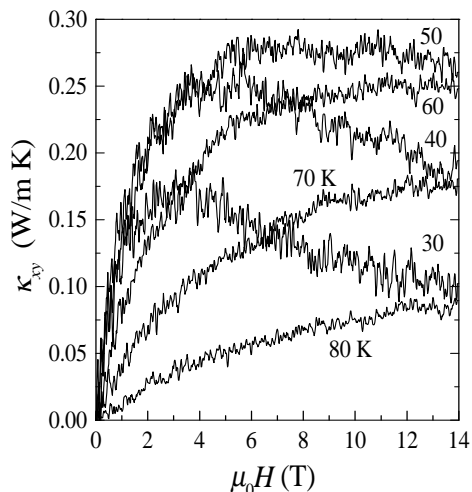,height=2.6in,width=2.7in}}
\caption{The Hall conductivity $\kappa_{xy}$ in untwinned $\rm YBa_2Cu_3O_7$ (sample 1), with 
thermal current ${\bf J}_Q \parallel {\bf a}$ and $\bf H\parallel c$. 
}
\label{F1}
\end{figure} 

Equation \ref{kxx} has been tested at high resolution. In underdoped YBCO (in which $p(T)$ 
approaches 4 T$^{-1}$ and a plateau value is observed below 15 K), $\kappa_{xx}$ fits Eq. 
\ref{kxx} closely over 2 decades in $H$ \cite{Krish3}.  These fits require the background 
term $\kappa_B$ to be $H$-independent \cite{Krish1,Krish2} (see also Yu {\em et al.} 
\cite{Fyu}).  Thus, these earlier findings impose the constraint that either the $qp$ 
conductivity or the phonon conductivity must be $H$-independent. 

\begin{figure}[h]
\centerline{\psfig{figure=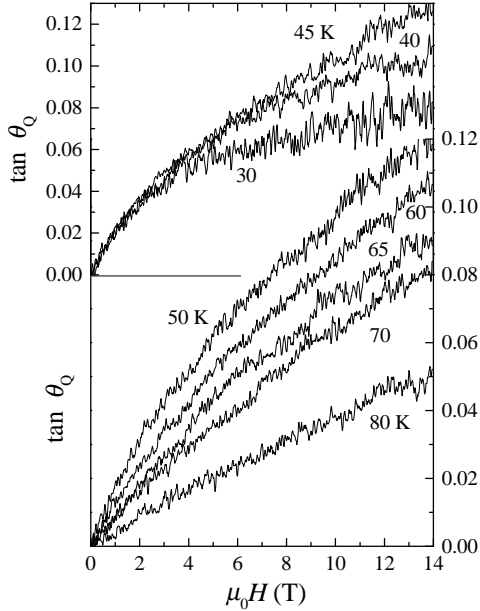,height=3.2in,width=2.8in}}
\caption{Traces of the Hall angle $\tan \theta_Q= \kappa_{xy}/\kappa_e$ where $\kappa_e\equiv 
\kappa_{xy}-\kappa_B$, with $\kappa_B$ determined by fitting to Eq. \ref{kxx} (sample 1).}
\label{F2}
\end{figure} 
\noindent

Because $\kappa_{xy}$ displays strong curvature vs. $H$, the phonon term is clearly the one 
that is $H$-independent.  Hence, in Eq. \ref{kxx}, we identify the difference $\kappa_{xx}-
\kappa_B$ with the field-dependent part of the $a$-axis $qp$ conductivity $\kappa_{e,a}(H,T)$ 
(we have written $\kappa_e^0 = \kappa_{e,a}(0,T)$).  The $H$-independent term $\kappa_B$ is 
identified with $\kappa_{ph}$, including possibly a very small residual electronic term 
$\kappa_e^{00}$ that is $H$-independent \cite{Franz}.  With $\kappa_{e,a}$ so extracted, we 
may obtain the thermal Hall angle $\tan\theta_Q\equiv \kappa_{xy}(H,T)/\kappa_{e,a}(H,T)$ 
(hereafter we drop the $a$ in $\kappa_{e,a}$ ).

The Hall conductivity $\kappa_{xy}$ in sample 1 (see Fig. \ref{F1}) displays behavior similar 
to that in Ref. \cite{Krish1}, but with stronger curvature vs. $H$. Results for samples 2 are 
closely similar.  Combining $\kappa_{xy}$ and the parameters in Eq. \ref{kxx}, we derive 
$\tan\theta_Q$.  The traces in Fig. \ref{F2} show that $\tan\theta_Q$ is initially linear in 
$H$, but displays increasing curvature at large $H$ and low $T$.  We focus only on the 
initial value.

\begin{figure}[h]
\centerline{\psfig{figure=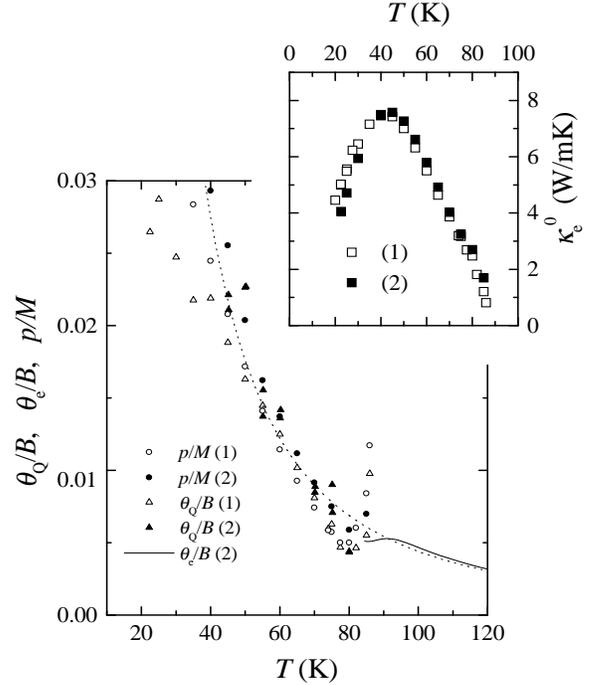,height=3.6in,width=3.1in}}
\caption{[Main panel] Comparison of the parameter $p(T)/{\cal M}$ in samples 1 (open circles) 
and 2 (solid circles), with the weak-field $\theta_{QP}/B$ in 1 (open triangles) and 2 (solid 
triangles).  The solid line are values of $\theta_e/B$ measured in sample 2 above 93 K (the 
broken line shows its extrapolation as the curve 44/$T^2$).  The inset shows the zero-field 
$qp$ conductivity $\kappa_e^0$ extracted in 1 and 2. }
\label{F3}
\end{figure}

The weak-field $\tan\theta_Q$ provides a powerful clue to the meaning of $p(T)$.  Comparing 
the two quantities (main panel of Fig. \ref{F3}), we note that $p(T)$ increases rapidly with 
decreasing $T$, closely tracking $\theta_Q$.  Between 30 and 80 K, the data for $p(T)$ and 
$\theta_Q$ (open triangles) may be made to lie on the same curve, if we simply divide $p(T)$ 
by the dimensionless scale factor ${\cal M}\equiv p(T)B/\theta_Q\; (B\rightarrow 0)$ (in both 
1 and 2, $\cal M$ equals 13).  Thus, the two quantities share the same $T$ dependence over a 
broad range of $T$.  

As $\tan\theta_Q$ is generally proportional to the $qp$ lifetime, the comparison implies that 
$p(T)$ is also proportional to the $qp$ lifetime or mean-free-path $\ell(T)$.  It is then 
convenient to express $p(T)$ in the general form 
\begin{equation}
p(T) = \ell(T)s/\phi_0,
\label{p}
\end{equation}
where the length scale $s$ depends on the particular model, and $\phi_0 = h/2e$ ($h$ is 
Planck's constant and $e$ is the elementary charge).  

\noindent

We may also express the initial Hall-angle in terms of a `Hall-angle' mean-free-path 
$\ell_H$, viz. $\theta_Q = \ell_H(T) eB/\hbar k_F$ with $k_F$ the Fermi wavevector.  
Proportionality between $p(T)$ and $\tan\theta_Q$ implies 
\begin{equation}
s = (\ell_H/\ell)({\cal M}\pi/k_F)\;\simeq 65(\ell_H/\ell)\;\rm \AA,
\label{s}
\end{equation}
with $k_F \simeq 0.64\;\rm\AA^{-1}$. 

\begin{figure}[h]
\centerline{\psfig{figure=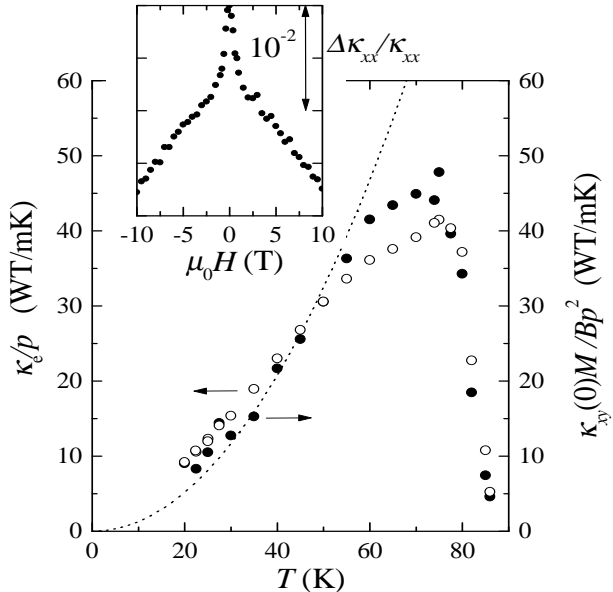,height=3.2in,width=3.2in}}
\caption{(Main panel) The $T$ dependence of $\kappa_e/p(T)$ (open circles) and 
$\kappa_{xy}(0)/Bp(T)^2$ (solid circles) in sample 1.  Both quantities are proportional to 
$c_e$ ($\kappa_{xy}(0)$ is multiplied by $\cal M$).  From the value of $\alpha$ reported by 
Wright {\em et al.}, we have calculated $\kappa_e^0/\ell_1$, and plotted it as 
$(\kappa_e^0/\ell_1) \times(\phi_0/s)$ with $s$ = 45 $\rm \AA$ (the broken line = 0.013 
$T^2\; \rm TW/Km$). The inset shows the cusp anomaly in $\kappa_{xx}$ vs. $H$ at 55 K in 
underdoped YBCO.
}
\label{F4}
\end{figure}

We add the caveat that, in 93-K YBCO, the simple proportionality between $p(T)$ and $\ell$ 
expressed in Eq. \ref{p} does not hold close to $T_c$. Above 80 K, the extracted $p(T)$ and 
$\theta_Q$ appear to diverge as $T\rightarrow T_c^-$.  This divergence is an artifact 
produced by the existence of a sharp zero-field cusp in the $\kappa_{xx}-H$ profile close to 
$T_c$ (its presence enhances the slope of $\kappa_{xx}$ vs. $H$ as $H\rightarrow 0$).  In 93-
K YBCO, the cusp smoothly merges into the regular profile and cannot be isolated easily.  
Thus, the apparent $p(T)$ is increasingly distorted as we get close to $T_c$.  However, in 
underdoped YBCO, the cusp is clearly distinct from the regular profile and its effect may be 
subtracted (see inset in Fig. \ref{F4}).  The origin of the cusp is unknown at present. 

Recent heat capacity experiments on YBCO place strong constraints on the analysis of 
$\kappa_e^0$ at low $T$, which we now address. The $qp$ thermal conductivity $\kappa_e^0$ 
($H$=0) extracted for samples 1 and 2 are displayed in the inset in Fig. \ref{F3}.  As $T$ 
decreases, $\kappa_e^0$ increases by $\sim$10 before decreasing rapidly.  We may write 
$\kappa_e^0 = c_e\langle v\ell\rangle/2$ where $c_e$ is the electronic heat capacity and 
$\langle v\ell\rangle$ denotes the group velocity-$mfp$ product averaged over the Fermi 
Surface.  The weak-field Hall conductivity is $(c_e\langle v\ell\rangle/2) \tan\theta_Q$.  At 
low $T$ in a $d$-wave superconductor, the Boltzmann-equation approach gives $\langle 
v\ell\rangle = (v_1\ell_1 + v_2\ell_2)/2$, where subscripts 1 and 2 refer to the principal 
axes of the Dirac cone at the nodes (as $v_1/v_2\simeq 7$, transport along the 1-axis 
dominates).  Dividing the two conductivities by $\ell_1$ and $\ell_1^2$, respectively, gives 
quantities that may be compared with $c_e$.  Although $p(T)$ in our experiment determines 
$\ell_1$ only up to the unknown parameter $s$, we may plot our data as $\kappa_e^0/p(T)$ and 
$\kappa_{xy}(0)/Bp(T)^2$, and examine their $T$ dependence (Fig. \ref{F4}).  Below 60 K, both 
are consistent with a power-law $T$ dependence (in particular, $\kappa_{xy}(0)/Bp(T)^2$ fits 
a $T^2$ behavior).

\noindent
The heat capacity measurements appear to be converging to the result that, at $H$=0, $c_e$ 
varies as $\alpha T^2$ at low $T$, with $\alpha \simeq 0.064\; {\rm mJK^{-3}mol^{-1}}$ 
\cite{Wright}.  This value of $\alpha$ implies \cite{ce} $v_1 = v_F = 1.78\times 10^7$ cm/s 
and $\kappa_e^0/\ell_1 = 2.72 \times10^4 T^2$.  By comparing the second equation with our 
results, we can fix the unknown scale $s$.  The best fit (broken line in Fig. \ref{F4}) gives 
$s= 45 \rm \AA$, in reasonable agreement with Eq. \ref{s}.  With this $s$ in Fig. \ref{F3}, 
we find that $\ell_1\simeq 1,800 \rm \AA$ at 40 K.  

We next compare the $T$ dependence of $\tan\theta_Q$ with the electrical Hall angle 
$\tan\theta_e = \rho_{xy}/\rho_b$ (the solid line in Fig. \ref{F3}, main panel).  Remarkably, 
$\tan\theta_Q$ falls on the $1/T^2$ curve of $\tan\theta_e$ extrapolated to below 93 K with 
no adjustments to either scale (dotted line).  In YBCO, a large fraction ($\sim 1/2$) of the 
charge carriers are in the chains ($\parallel \bf b$).  Because of the resistivity anisotropy 
$\rho_a/\rho_b$ (=2.1 at 100 K), it is important that we compare $\tan\theta_Q$ with the 
correct Hall angle $\rho_{xy}/\rho_b$ (this probes the electrical current $\parallel {\bf 
a}$; the incorrect choice $\rho_{xy}/\rho_a$ is only half as large).  Hence, our finding of 
numerical agreement between $\tan\theta_Q$ and $\tan\theta_e$ applies strictly to the 
carriers in the $\rm CuO_2$ plane.  The case for ${\bf J}_Q\parallel \bf b$ remains to be 
investigated.  Nevertheless, when twinned crystals are used (see 3 and 4), we still observe 
{\em equality} between the two Hall angles.  This suggests that the equality holds as well 
when ${\bf J}_Q\parallel \bf b$.  Thus, our overall finding is inconsistent with that of 
Zeini {\em et al.} who find that $\theta_Q/\theta_e\simeq 2$ in a twinnned crystal 
\cite{Zeini}.

\begin{figure}[h]
\centerline{\psfig{figure=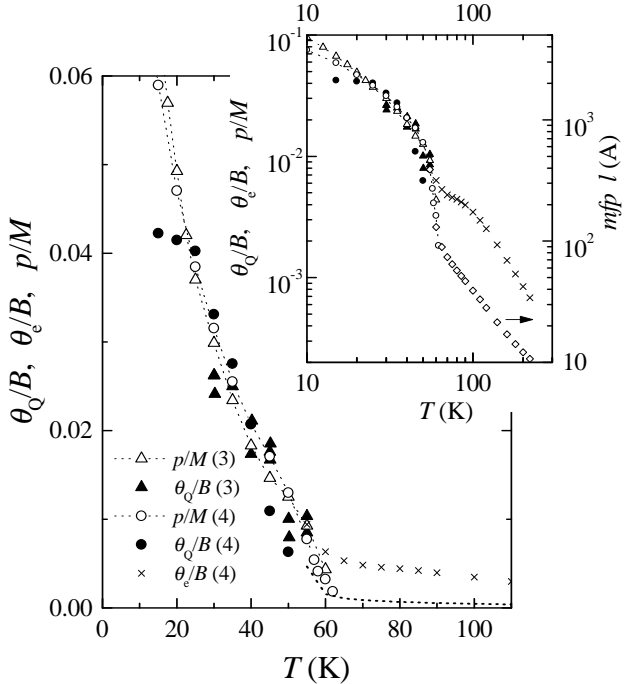,height=3.6in,width=3.3in}}
\caption{[Main panel] The $T$ dependence of $p(T)/{\cal M}$ (open symbols) and $\theta_Q/B$ 
(solid symbols) in underdoped YBCO (samples 3 and 4).  Data for $\theta_e/B$ above 60 K (from 
4) are shown as an $\times$.  The inset shows the same data in log-log scale to emphasize 
that $\theta_Q$ is continuous with $\theta_e$ except for a `dip' between 60 and 100 K 
associated with the opening of the pseudogap.  The right axis represents the data in terms of 
$\ell$ and $\ell_H$ (see text).  The diamonds are values for $\ell$ obtained from the in-
plane $\rho$ measured in 4 (also shown as a dotted line in main panel).}
\label{F5}
\end{figure}

The results in the underdoped 60-K crystals provide important complementary information.  We 
may extend measurements of $p(T)$ to lower $T$ (10 K) because flux pinning is much weaker.  
Further, the cusp near $T_c$ is readily separated from the regular $H$ dependence of 
$\kappa_{xx}$ (inset of Fig. \ref{F4}), so $p(T)$ may be reliably determined up to $T_c$.  In 
the main panel of Fig. \ref{F5}, we display $p(T)$ measured in samples 3 and 4 (open 
triangles and circles, respectively).  With increasing $T$, $p(T)$ falls rapidly.  The $T$ 
dependence of $\theta_Q$ (solid symbols) is closely similar.  By adjusting $\cal M$ (= 20 and 
15 for samples 3 and 4, respectively), we may also match the curves for $\theta_Q$ and 
$p(T)$.  We emphasize that $p(T)$ does not show a divergence near $T_c$, unlike the case in 1 
and 2.  Moreover, as $T$ increases above 55 K, $p(T)/M$ falls {\em below} the curve for 
$\theta_e/B$.  Also shown are data for the electrical Hall angle $\theta_e$ measured in 4 
(crosses). 

In the inset in Fig. \ref{F5}, we display the same data over a broader range of $T$ (10 to 
240 K).  The log-log plot shows that $\theta_Q$ continues on the curve for $\theta_e$ 
extrapolated below $T_c$ (without any scale adjustment).  Between $T_c$ and 100 K, the upper 
curve displays an apparent broad minimum.  This shallow feature is the superposition 
of a rapidly decreasing $\tan\theta_e$ (as $T\rightarrow T_c^+$) and a rapidly increasing 
$\tan\theta_Q$ as $T$ decreases below $T_c$.  A recent investigation reveals that the 
decrease in $\tan\theta_e$ is correlated with the opening of the pseudogap at 150 K 
\cite{Xu}.  This causes the Hall angle to plummet steeply relative to its high-$T$ trend.  
The inset shows rather clearly that, apart from the dip, $\theta_Q$ is continuous with the 
normal-state $\theta_e$.  

It is instructive to re-express the underdoped results in terms of mean-free-paths.  Over the 
interval 30-55 K, $\ell$ and $\ell_H$ are proportional.  If we assume $\ell\simeq\ell_H$, we 
may use the scale factor $\cal M$ to convert $p(T)$ and $\theta_Q$ into $mfp$'s.  In the 
inset in Fig. \ref{F5} (right axis) we have expressed $p(T)$ and $\theta_Q$ in terms of 
$\ell$.  We find that $\ell$ decreases rapidly from 4,000 to 550 $\rm \AA$ betweem 10 and 50 
K.  Above $T_c$, however, $\ell$ and $\ell_H$ follow different curves.  Whereas $\ell_H$ 
smoothly joins the $1/T^2$ Hall-angle curve in the normal state, $\ell$ derived from $p(T)$ 
sharply decreases across $T_c$ (open symbols), assuming values above 55 K much shorter than 
$\ell_H$.  The sharp reduction in $\ell$ is consistent with the transport $mfp$ (diamonds) 
deduced from the in-plane resistivity $\rho$ \cite{rho}.  

\noindent
It appears that, starting about 10 K below $T_c$, a novel scattering mechanism $\Gamma_{tr}$ 
becomes prominent and grows rapidly with $T$.  It selectively damps the longitudinal current, 
but does not seem to affect the transverse (Hall) current.  Thus, $\ell$ decreases more 
rapidly than $\ell_H$ across $T_c$, resulting in the two branches shown in the inset of Fig. 
\ref{F5}.  The two branches recall the two-lifetime scenario in the Hall-angle experiment of 
Chien {\em et al.} \cite{Chien}. 

In summary, we have combined measurements of $\kappa_{xx}$ vs. $H$ with $\kappa_{xy}$ to 
achieve a self-consistent separation of the $qp$ and phonon conductivities.  The numerical 
agreement between the thermal and electrical Hall angles in both twinned and untwinned 
crystals, and the comparison with heat capacity experiments support the validity of the 
analysis presented.

We thank P. W. Anderson, M. Franz, and F.D.M. Haldane for many discussions.  N.P.O. 
acknowledges support from the U.S. Office of Naval Research (N00014-90-J-1013) and the U.S. 
National Science Foundation (DMR98-09483).
\vskip2mm\noindent
$^\dagger$ {\em Permanent address of ZAX: Department of Physics, Zhejiang University, 
Hangzhou, China.}\newline
$^\S$ {\em Present address of LT: Department of Physics, University of Toronto, Toronto, 
Ontario, Canada M5S 1A7.}


%
%
%
%

\begin{references}
\bibitem{ce}  At low $T$, $c_e$ is given by $4\eta k_B^3T^2/(\pi d\hbar^2 v_Fv_2)$, where $d 
= 5.8\;\rm \AA$, $v_f$ and $v_2$ are the principal-axis velocities, and $\eta= \int_0^\infty 
dx\; x^3(-\partial f/\partial x)\simeq 5.41$.  With $v_F/v_2 = 7$, we find $v_F=1.78\times 
10^7$ cm/s.
\bibitem{rho} From the measured $\rho$ (70 $\mu{\rm \Omega}$cm at 70 K), we find 
$k_F\langle\ell\rangle = 22$, which gives a lower bound to the $mfp$ of 34 $\rm \AA$.  We 
estimate that near the nodes $\ell$ is about 3 times larger.
\end{references}
\end{document}